\documentclass[12pt]{article}
\usepackage{amssymb}
\usepackage{graphicx}
\usepackage[cp1251]{inputenc}
\usepackage{rotating}

\begin{document}
 \noindent {\footnotesize\it Astronomy Letters, 2023, Vol. 49, No. 9, pp. 493--500.}
 \newcommand{\dif}{\textrm{d}}

 \noindent
 \begin{tabular}{llllllllllllllllllllllllllllllllllllllllllllll}
 & & & & & & & & & & & & & & & & & & & & & & & & & & & & & & & & & & & & & &\\\hline\hline
 \end{tabular}

  \vskip 0.5cm
  \bigskip
 \bigskip
 \centerline{\large\bf Estimation of the Galactocentric Distance of the Sun from}
 \centerline{\large\bf Cepheids Close to the Solar Circle}

 \bigskip
 \bigskip
  \centerline { 
   V. V. Bobylev\footnote [1]{vbobylev@gaoran.ru}}
 \bigskip
 \centerline{\small\it Pulkovo Astronomical Observatory, Russian Academy of Sciences, St. Petersburg, 196140 Russia}
 \bigskip
 \bigskip
{Based on Cepheids located near the solar circle, we have determined the Galactocentric distance of the Sun $R_0$ and the Galactic rotation velocity at the solar distance $V_0$. For our analysis we used a sample of $\sim$200 classical Cepheids from the catalogue by Skowron et al. (2019), where the distances to
them were determined from the period--luminosity relation. For these stars the proper motions and line-of-sight velocities were taken from the Gaia~DR3 catalogue. The values of $R_0$ found lie within the range [7.8--8.3] kpc, depending on the heliocentric distance of the sample stars, on the adopted solar velocity
relative to the local standard of rest, and on whether or not the perturbations caused by the Galactic spiral density wave are taken into account. The dispersion of the $R_0$ estimates is $\sim$2~kpc. Similarly, the values of $V_0$ lie within the range [240--270]~km s$^{-1}$ with a dispersion of the estimates of 70--90~km s$^{-1}$. We consider the following estimates to be the final ones: $R_0=8.24\pm0.20$~kpc and $V_0=268\pm8$~km s$^{-1}$ found by taking into account the perturbations from the Galactic spiral density wave.
 }

\bigskip
\section*{INTRODUCTION}
The Galactocentric distance of the Sun $R_0$ is an important parameter for studying the structure, kinematics, and dynamics of the Galaxy. Various methods
of estimating this quantity are known. According to
Reid (1993), such measurements can be divided into
direct, secondary, and indirect. Bland-Hawthorn and
Gerhard (2016) propose to divide such measurements
into direct, model-dependent, and secondary. Nikiforov (2004) proposed a special classification with the division of measurements into three classes, depending on the type of measurements, the type of $R_0$ estimates, and the type of reference objects.

A truly direct method is to determine the absolute
trigonometric parallax of an object located near
the Galactic center. Based on VLBI observations
of several maser sources in the region Sgr~B2, Reid
et al. (2009) obtained an estimate of $R_0=7.9^{+0.8}_{-0.7}$~kpc
by this method. The dynamical parallax method can
also be assigned to the direct ones. At present, an
analysis of the orbital motion of stars around the
central supermassive black hole allows $R_0$ to be estimated
by this method with a relative error of about 0.3\% (Abuter et al. 2019). According to one of the latest determinations of the GRAVITY Collaboration,
$R_0=8275\pm9_{\rm stat.} \pm33_{\rm sys.}$~pc (Abuter et al. 2021).

Such variable stars as classical Cepheids, type II Cepheids, and RR Lyrae variables are important for estimating $R_0$. A high accuracy of the distance estimates
for these variables is possible owing to the
period--luminosity relation (Leavitt 1908; Leavitt and Pickering 1912) and the period--Wesenheit relation (Madore 1982), which have currently been well calibrated using highly accurate trigonometric parallaxes of stars (Ripepi et al. 2019).

Type II Cepheids and RR Lyrae variables are distributed over the entire Galaxy. Relatively younger classical Cepheids are distributed over the entire
Galactic disk. Their geometric and kinematic properties
are used to estimate $R_0$. For example, the
R0 estimation methods based on the assumption
about a symmetric distribution of Cepheids in the
Galactic bulge or a symmetric distribution of globular
cluster member variable stars in the Galaxy
are known. The estimates of $R_0$ are quite often
obtained by analyzing the Galactic rotation curve,
where $R_0$ acts as a sought-for unknown, along with
other parameters. However, the estimates of $R_0$ are
obtained by analyzing the Galactic rotation curve not
only from Cepheids, but also from other Galactic disk
objects~--- masers, OB stars, open star clusters, etc.

In addition to the first-class distances to Cepheids, it is important to have their highly accurate proper motions and line-of-sight velocities. From this point
of view, of great interest is the Gaia space experiment (Prusti et al. 2016), which is devoted to the determination of highly accurate trigonometric parallaxes,
proper motions, and a number of photometric characteristics
for more than 1.5 billion stars. In the recently
published GaiaDR3 version (Vallenari et al. 2022),
the line-of-sight velocities of stars were improved
significantly~--- the previously measured values were
refined and the new ones were determined for a large
number of stars. At the same time, the trigonometric
parallaxes and proper motions of stars were copied
from the previous version of the Gaia~EDR3 catalogue
(Gaia Early Data Release~3, Brown et al. 2021).

Various authors regularly perform reviews of individual $R_0$ estimates with the derivation of their mean. For example, Nikiforov (2004) found $R_0=7.9\pm0.2$~kpc based on 65 original measurements from 1974 to 2004, Malkin (2013) calculated the
mean $R_0=8.0\pm0.25$~kpc by analyzing 53 individual estimates from 1992 to 2011, and Bobylev and Bajkova (2021) found $R_0=8.1\pm0.1$~kpc based on 56 measurements from 2011 to 2021.

The goal of this paper is to estimate $R_0$ and the circular rotation velocity of the Galaxy at the solar distance $V_0$ using a large sample of classical
Cepheids located near the solar circle. For this
purpose, we use Cepheids from the catalogue by
Skowron et al. (2019), where the distances to them
were determined from the period--luminosity relation
with a mean error of about 5\%. We took the proper
motions and line-of-sight velocities of these Cepheids from the Gaia~DR3 catalogue. We applied a peculiar method of estimating $R_0$ and $V_0$ based on the analysis of objects close to the solar circle. The method is interesting in that $R_0$ and $V_0$ can be estimated even from one star located on the solar circle.

 \section*{METHOD}
To estimate the Galactocentric distance of the Sun $R_0$ and the circular rotation velocity of the Galaxy at the solar distance $V_0$ using objects located near the
solar circle, we know a method (see, e.g., Schechter et al. 1992) that we will call the classic one:
 \begin{equation}
 R_0={r\over 2\cos l},\quad V_0=-{V_l\over 2\cos l},
 \label{R-Vp-0}
 \end{equation}
where $r$ is the heliocentric distance of the star and $V_l$ is the velocity directed along the Galactic longitude ($V_l=4.74r\mu_l\cos b$). Here, the line-of-sight velocity $V_r$ of each star lying on the solar circle is assumed to be zero.

Sofue et al. (2011) proposed the following modification of the classic method: 
\begin{equation}
 R_0={r\over 2\cos l}(1-d/r),
\label{R-0}
 \end{equation}
\begin{equation}
 V_0=-{V_l\over 2\cos l}(1-d/r)+V_r\cot l,
\label{V-0}
 \end{equation}
where $d$ is the distance along the line of sight to the star from the solar circle,
\begin{equation}
 d=-{V_r\over A\sin 2l},
\label{dd-0}
 \end{equation}
$A$ is the Oort constant whose value in this paper is taken to be $A=15$~km s$^{-1}$ kpc$^{-1}$. The velocities $V_r$ and $V_l$ should be given relative to the local standard of rest. To reduce the observed heliocentric velocities of the stars to the local standard of rest, we use two sets of velocities. In the first case, we take the velocities from Sch\"onrich et al. (2010):
 \begin{equation}
 \label{V_LSR-Shoenrich}
 (U,V,W)_\odot=(11.10,12.24,7.25)~\hbox{km s$^{-1}$},
  \end{equation}
while in the second case we use the velocities found from a large sample of classical Cepheids in Bobylev and Bajkova (2023):
 \begin{equation}
 \label{V_LSR-RAA}
 (U,V,W)_\odot=(9.39,15.96,6.88)~\hbox{km s$^{-1}$}.
  \end{equation}
The velocities (6) are close to the corresponding velocities of the so-called standard solar apex  $(U,V,W)_\odot=(10.3,15.3,7.7)$~km s$^{-1}$ that were used
by Sofue et al. (2011) to analyze the masers Onsala~1 and Onsala~2N. When we repeated their calculations, it was found that the estimate of $R_0$ depends fairly
strongly on the adopted solar velocity relative to the local standard of rest. Therefore, in this paper we use both sets of such velocities.

The errors in $R_0$ and $V_0$ are estimated in accordance
with the following formulas:
\begin{equation}
 \delta R_0={1\over 2\cos l}
 \left[\delta r^2+ \left({\delta V_r\over A\sin 2l}\right)^2 \right]^{1/2},
\label{errR-0}
 \end{equation}
\begin{equation}
 \renewcommand{\arraystretch}{2.4}
 \begin{array}{lll}
 \displaystyle
 \delta V_0={1\over 2\cos l}
 \left[\delta V_l^2+ V_l^2\delta {V_r}^2 \left({1\over Ar\sin 2l}
 -{2\cos^2 l\over V_l\sin l}\right)^2  \right]^{1/2}.
\label{errV-0}
 \end{array}
 \end{equation}
According to the approach of Sofue et al. (2011), the
fact that a star belongs to the region of the solar circle
should be reconciled with its observed velocities and
the parameters of the Galactic rotation curve.

The errors (7) and (8) are calculated for each star; in what follows, they can serve as weights when calculating the weighted mean, as was done in Sofue et al. (2011). In this paper we prefer to calculate the mean values of $R_0$ and $V_0$, the standard deviations $\sigma_{R_0}$ and $\sigma_{V_0},$, and the errors of the mean $\varepsilon_{R_0}$ and $\varepsilon_{V_0}$ using the well--known relations.

Expressions (2) and (3) were derived by Sofue et al. (2011) under the assumption of purely circular motions of stars around the Galactic center. In Bobylev (2013) we proposed a modification of the method with the elimination of the systematic noncircular motions of stars related to the influence of the Galactic spiral density wave. The following formulas serve to take into account these effects:
 \begin{equation}
 \renewcommand{\arraystretch}{1.0}
 \begin{array}{lll}
 V_r=-u_\odot\cos b\cos l-v_\odot\cos b\sin-w_\odot\sin b+f_r(GR) \\
    +\tilde{v}_\theta\sin(l+\theta)\cos b
    -\tilde{v}_R \cos(l+\theta)\cos b+ V',
 \label{EQ-1}
 \end{array}
 \end{equation}
 \begin{equation}
 \begin{array}{lll}
 V_l=u_\odot\sin l-v_\odot\cos l+f_p(GR)
  +\tilde{v}_\theta \cos(l+\theta)+\tilde{v}_R\sin(l+\theta)
  + V',
 \label{EQ-2}
 \end{array}
 \end{equation}
where $(u_\odot,v_\odot,w_\odot)$ is the group velocity of the stars under consideration caused by the peculiar solar motion, the functions describing the differential Galactic rotation, whose specific form in our case is unimportant, are denoted by $f_r(GR)$ and $f_p(GR)$, and the influence of the residual effects is denoted by $V'$.

To take into account the influence of the spiral
density wave, we used a kinematic model based on
the linear density wave theory of Lin and Shu (1964),
in which the potential perturbation has the form of a
traveling wave. Then,
 \begin{equation}
 \begin{array}{rll}
      \tilde{v}_R=f_R \cos \chi,\\
 \tilde{v}_\theta=f_\theta \sin \chi,
 \label{VR-Vtheta}
 \end{array}
 \end{equation}
where $f_R$ and $f_\theta$ are the perturbation amplitudes of the radial (the perturbation is directed to the Galactic center in the spiral arm) and tangential (directed along the Galactic rotation) velocities; the phase of the wave is
 \begin{equation}
   \chi=m[\cot (i)\ln (R/R_0)-\theta]+\chi_\odot,
 \label{chi-creze}
 \end{equation}
where $i$ is the pitch angle of the spirals ($i<0$ for winding spirals); $m$ is the number of spiral arms; $\theta$ is the position angle of the star: $\tan\theta=y/(R_0-x)$, where $x, y$ are the heliocentric Galactic rectangular
coordinates of the star, with the $x$ axis being directed
from the Sun to the Galactic center and the direction
of the $y$ axis being coincident with the direction of
Galactic rotation; $\chi_\odot$ is the phase angle of the Sun.
The spiral wavelength $\lambda$, i.e., the distance (along
the Galactocentric radial direction) between the adjacent
spiral arm segments in the solar neighborhood,
is calculated from the relation tan $\tan |i|=\lambda m/(2\pi R_0)$.
Therefore, in our modeling we can specify either the
pitch angle or the wavelength.

\begin{figure}[t]
{ \begin{center}
  \includegraphics[width=0.65\textwidth]{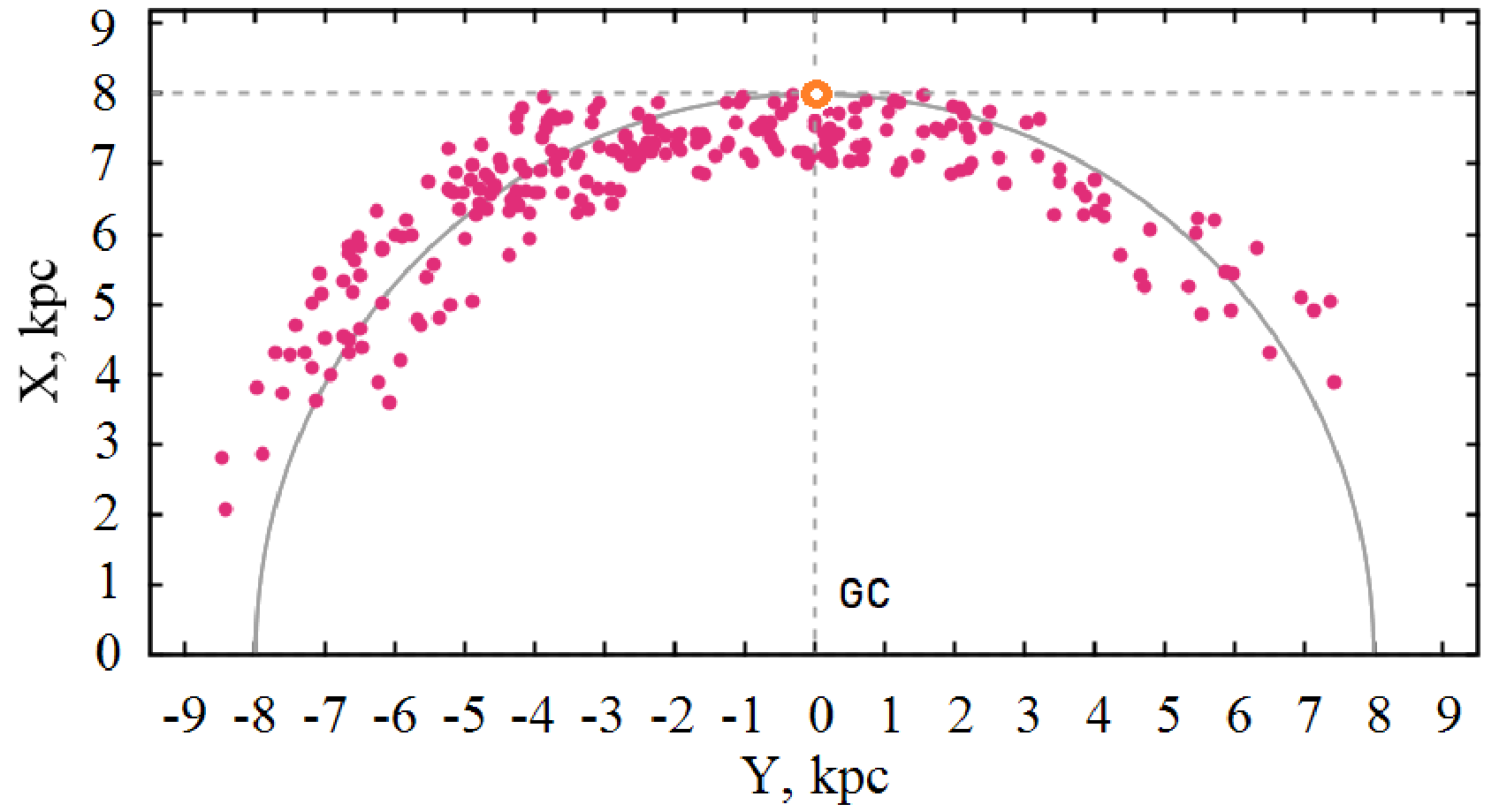}
  \caption{
Distribution of the Cepheids selected for our analysis in projection onto the Galactic $XY$ plane, the circle of radius 8 kpc is shown, the position of the Sun is marked by the yellow circle, GC is the Galactic center.}
 \label{XY}
\end{center}}
\end{figure}
\begin{figure}[t]
{ \begin{center}
  \includegraphics[width=0.9\textwidth]{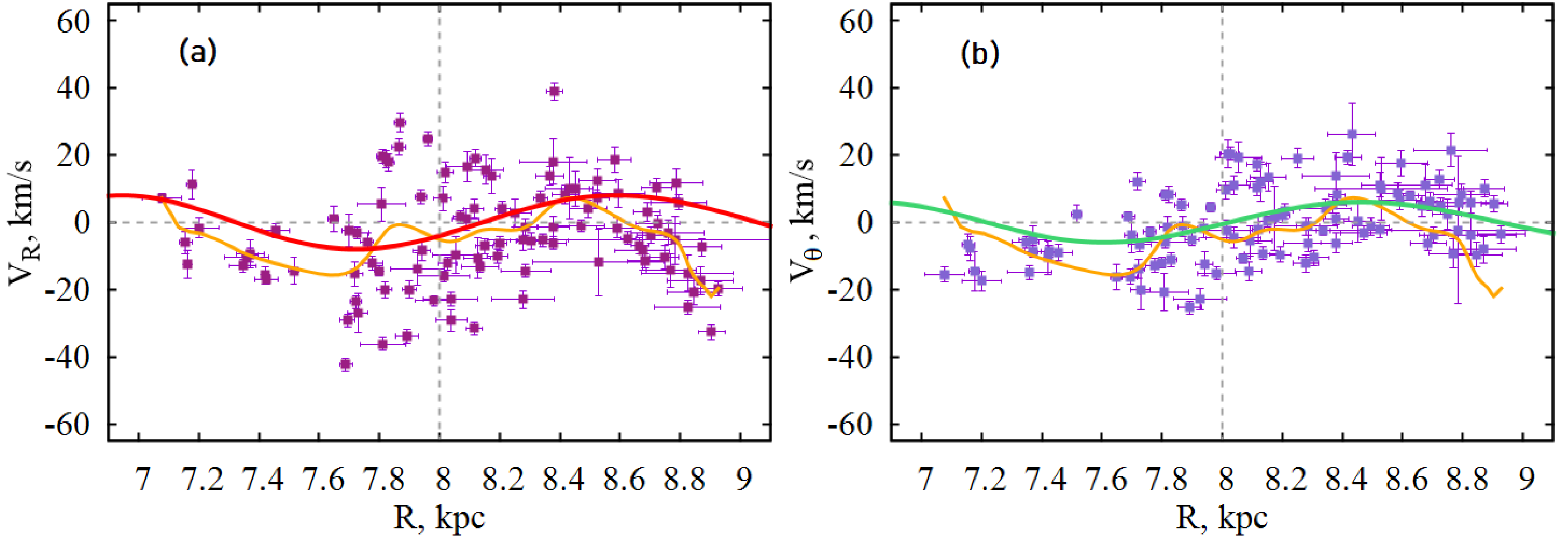}
  \caption{
Radial velocities of the Cepheids $V_R$~(a) and their tangential velocities $V_\theta$~(b) as a function of distance $R$.}
 \label{f-waves}
\end{center}}
\end{figure}

  \begin{table}[t]
  \caption[]{\small 
The estimates of $R_0$ and $V_0$ obtained by the classic method (1)
 }
  \begin{center}  \label{Table-1}    \small
  \begin{tabular}{|l|c|c|c|c|c|c|}\hline
 Parameters & $r>2$~kpc & $r>2.5$~kpc & $r>3.0$~kpc & $r>3.5$~kpc & $r>4$~kpc \\\hline

$N_{\star}$ & 117  &  104   & 94  & 84  & 75 \\
$R_0\pm\varepsilon_{R_0},$~kpc  &$7.68\pm0.22$&$7.98\pm0.22$&$8.24\pm0.21$&$8.32\pm0.18$&$8.41\pm0.19$\\
$\sigma_{R_0},$~kpc & 2.39 & 2.20 & 2.02 & 1.69 & 1.61 \\

$V_0\pm\varepsilon_{V_0},$~km s$^{-1}$ &$248\pm8$&$259\pm8$&$265\pm8$&$270\pm8$&$271\pm8$\\
$\sigma_{V_0},$~km s$^{-1}$ & 87 & 82 & 79 & 71 & 67 \\
 \hline
 \end{tabular}\end{center}
 \end{table}
  \begin{table}[t]
  \caption[]{\small
The estimates of $R_0$ and $V_0$ obtained by the method of Sofue et al. (2011) using relations (2)--(8)}
  \begin{center}  \label{Table-2}    \small
  \begin{tabular}{|l|c|c|c|c|c|c|}\hline
Parameters & $r>2$~kpc & $r>2.5$~kpc & $r>3.0$~kpc & $r>3.5$~kpc & $r>4$~kpc\\\hline

$N_{\star}$ & 119 &  105 & 94  & 86 & 78 \\
$R_0\pm\varepsilon_{R_0},$~kpc  &$7.44\pm0.21$&$7.73\pm0.21$&$7.83\pm0.21$&$7.89\pm0.22$&$7.97\pm0.23$\\
$\sigma_{R_0},$~kpc & 2.28 & 2.15 & 2.08 & 2.06  & 2.07  \\

$V_0\pm\varepsilon_{V_0},$~km s$^{-1}$ &$239\pm8$&$249\pm8$&$249\pm8$&$252\pm8$&$252\pm8$\\
$\sigma_{V_0},$~km s$^{-1}$ & 82 & 79 & 76 & 74 & 74 \\
 \hline
 \end{tabular}\end{center}
 \end{table}
  \begin{table}[t]
  \caption[]{\small
The estimates of $R_0$ and $V_0$ obtained by the method of Sofue et al. (2011) by additionally taking into account the influence of the spiral density wave with the following parameters: $m=4$, $\lambda=1.7$~kpc, $\chi_\odot=-60^\circ,$ $f_R=7$~km s$^{-1}$ and $f_\theta=4$~km s$^{-1}$
}
  \begin{center}  \label{Table-3}    \small
  \begin{tabular}{|l|c|c|c|c|c|c|}\hline
Parameters & $r>2$~kpc & $r>2.5$~kpc & $r>3.0$~kpc & $r>3.5$~kpc & $r>4$~kpc\\\hline

$N_{\star}$ & 122 &  109 & 97 & 89 & 80 \\
$R_0\pm\varepsilon_{R_0},$~kpc  &$7.38\pm0.21$&$7.67\pm0.20$&$7.78\pm0.21$&$7.78\pm0.22$&$7.96\pm0.22$\\
$\sigma_{R_0},$~kpc & 2.32 & 2.14 & 2.06 & 2.06 & 1.99 \\

$V_0\pm\varepsilon_{V_0},$~km s$^{-1}$ &$244\pm8$&$253\pm8$&$252\pm8$&$254\pm8$&$257\pm9$\\
$\sigma_{V_0},$~km s$^{-1}$ & 87 & 82 & 77 & 78 & 77 \\
 \hline
 \end{tabular}\end{center}
 \end{table}

 \section*{DATA}
The paper by Skowron et al. (2019), where the distances, ages, pulsation periods, and photometric data are given for 2431 classical Cepheids, served us as the basis for our study. The observations of these variable stars were performed within the OGLE (Optical Gravitational Lensing Experiment) program (Udalski et al. 2015). The distances to the Cepheids were calculated based on the calibration period--luminosity relations found by Wang et al. (2018) from the Cepheid light curves in the mid--infrared for eight bands. These include four bands of the WISE (Widefield
Infrared Survey Explorer) catalogue (Chen et al. 2018), W1--W4: [3.35], [4.60], [11.56], and [22.09] $\mu$m, and four bands of the GLIMPSE (Spitzer
Galactic Legacy Infrared Mid--Plane Survey Extraordinaire)
survey (Benjamin et al. 2003): [3.6], [4.5], [5.8], and [8.0] $\mu$m. In the catalogue by Skowron et al. (2019) the extinction $A_{K_s}$ was calculated for each star from extinction maps. According to these authors, the Cepheid distance error in their catalogue is $\sim$5\%. Skowron et al. (2019) estimated the ages
using the technique of Anderson et al. (2016) by taking into account the stellar rotation period and the metallicity.
 
In Bobylev et al. (2021), while studying the kinematics
of a large sample of classical Cepheids, we
showed that the scale of Skowron et al. (2019) should
be slightly extended. Therefore, in this paper the
distances to the Cepheids from this catalog were increased
by 10\%. For our analysis we select Cepheids
from the $R$ range 7--9 kpc by calculating the Galactocentric
distance $R$ for them using a preliminary value
of $R_0=8.0$~kpc. 
 
In the method of Sofue et al. (2011) using relations
(2)--(8), it is required that the stars lie in the
first and fourth Galactic quadrants, i.e., in the fairly
narrow longitude range $|l|<80^\circ$, and it is necessary
to have $d\ll r$. Therefore, we apply a constraint on
the coordinate x and perform our calculations with
various constraints on the distance $r$. To improve the homogeneity of the sample, we take Cepheids younger than 120 Myr. Finally, there should be no high line-of-sight velocities. As a result, we use the following constraints:
 \begin{equation}
 \begin{array}{lll}
 7<R<9~\hbox {kpc},\\
 x>0~\hbox {kpc},\\
 \hbox {AGE}<120~\hbox {Myr},\\
 |V_r|<20~\hbox {km s$^{-1}$}.
 \label{otbor}
 \end{array}
 \end{equation}
Under such selection conditions we obtained a sample of about 200 Cepheids with a mean age of 87 Myr. Figure 1 presents the distribution of the selected
Cepheids in projection onto the Galactic $XY$ plane,
where the $X$ axis is directed from the Galactic center
toward the Sun and the $Y$ axis is directed along the Galactic rotation.

In Fig. 2 the radial, $V_R$, and tangential, $V_\theta$, velocities
of the Cepheids are plotted against the distance
$R$. Here, the Cepheids were taken under the
condition $r>3$~kpc. A periodicity in both radial and
tangential velocities of the Cepheids is clearly seen.
The yellow lines in the figure indicate the averaged velocities (averaged by the GNUPLOT code), while the thick red and green periodic curves indicate the
influence of the spiral density wave chosen by us. For
this purpose, we used a four-armed ($m=4$) Galactic spiral pattern with a wavelength $\lambda=1.7$~kpc and velocity perturbation amplitudes $f_R$ and $f_\theta$ of 7 and
4 km s$^{-1}$, respectively.

 \section*{RESULTS AND DISCUSSION}
Tables 1--3 give the values of $R_0$ and $V_0$ obtained
by three methods both without and with allowance for
the influence of the spiral density wave. The data in
these tables were calculated using the parameters of
the solar motion relative to the local standard of rest (5).

  \begin{table}[t]
  \caption[]{\small
The estimates of $R_0$ and $V_0$ obtained by the method of Sofue et al. (2011) using relations (2)--(8) and the parameters of the solar motion relative to the local standard of rest (6)
 }
  \begin{center}  \label{Table-4}    \small
  \begin{tabular}{|l|c|c|c|c|c|c|}\hline
Parameters & $r>2$~kpc & $r>2.5$~kpc & $r>3.0$~kpc & $r>3.5$~kpc & $r>4$~kpc\\\hline

$N_{\star}$ & 107 &  96 & 89  & 82 & 75 \\
$R_0\pm\varepsilon_{R_0},$~kpc  &$8.07\pm0.22$&$8.30\pm0.22$&$8.30\pm0.22$&$8.23\pm0.22$&$8.31\pm0.22$\\
$\sigma_{R_0},$~kpc & 2.31 & 2.19 & 2.03 & 1.97 & 1.93 \\

$V_0\pm\varepsilon_{V_0},$~km s$^{-1}$ &$263\pm8$&$269\pm8$&$265\pm8$&$265\pm8$&$265\pm8$\\
$\sigma_{V_0},$~km s$^{-1}$ & 87 & 84 & 77 & 75 & 73 \\
 \hline
 \end{tabular}\end{center}
 \end{table}
  \begin{table}[t]
  \caption[]{\small
The estimates of $R_0$ and $V_0$ obtained by the method of Sofue et al. (2011) with the parameters of the solar motion relative to the local standard of rest (6) by additionally taking into account the influence of the spiral density wave with
the following parameters: $m=4$, $\lambda=1.7$~kpc, $\chi_\odot=-60^\circ,$ $f_R=7$~km s$^{-1}$ and $f_\theta=4$~km s$^{-1}$
 }
  \begin{center}  \label{Table-5}    \small
  \begin{tabular}{|l|c|c|c|c|c|c|}\hline
Parameters & $r>2$~kpc & $r>2.5$~kpc & $r>3.0$~kpc & $r>3.5$~kpc & $r>4$~kpc\\\hline

$N_{\star}$ & 114 &  102 & 93 & 84 & 76 \\
$R_0\pm\varepsilon_{R_0},$~kpc  &$7.80\pm0.21$&$8.14\pm0.20$&$8.24\pm0.20$&$8.27\pm0.20$&$8.35\pm0.21$\\
$\sigma_{R_0},$~kpc & 2.21 & 2.04 & 1.91 & 1.84 & 1.82 \\

$V_0\pm\varepsilon_{V_0},$~km s$^{-1}$ &$258\pm8$&$268\pm8$&$268\pm8$&$269\pm8$&$269\pm9$\\
$\sigma_{V_0},$~km s$^{-1}$ & 89 & 85 & 80 & 78 & 78 \\
 \hline
 \end{tabular}\end{center}
 \end{table}

Tables 4 and 5 give the values of $R_0$ and $V_0$ obtained by two methods both without and with allowance for the influence of the spiral density wave
(just as in Tables 2 and 3) with the parameters of the
solar motion relative to the local standard of rest (6).

As can be seen from Table 1, the values of $R_0$ and $V_0$ found by the classic method agree satisfactorily with the known ones, for example, those
pointed out in the Introduction. However, the classic
method has an important shortcoming---the line-of-sight
velocities of all the Cepheids under consideration
are nonzero. Therefore, the results presented
in Tables 2--5 are more interesting, since they were
obtained by a method free from this shortcoming.
Allowance for the perturbations from the spiral
density wave has a favorable effect on the estimates
of $R_0$ and $V_0$. The values of $R_0$ and $V_0$ do not
change fundamentally, but the dispersions $\sigma_{R_0}$ and
the errors $\varepsilon_{R_0}$ decrease, though slightly. Therefore,
we give preference to the parameters given in Tables 3 and 5. We are oriented to the values obtained under the constraint $r>3.0$~kpc, since there are still quite a
few stars here and the errors are small.

Note that the main interest in this paper is related to the determination of $R_0$, while the Galactic rotation velocity at the solar distance $V_0$ is determined more
reliably by analyzing the kinematics of large stellar
samples. For example, Mr\'oz et al. (2021) found
$V_0=233.6\pm2.8$~km s$^{-1}$ from the kinematics of 773
classical Cepheids with proper motions and line-of-sight
velocities from the Gaia~DR2 catalogue (Brown
et al. 2018); Bobylev et al. (2021) obtained an estimate
of $V_0=240\pm3$~km s$^{-1}$ from the kinematics
of $\sim$800 classical Cepheids with proper motions and
line-of-sight velocities from the Gaia~DR2 catalogue; Eilers et al. (2021) found 
$V_0=229.0\pm0.2$~km s$^{-1}$ from a sample of $\sim$23\,000 red giants.

As can be seen from our comparison of the data
in Tables 2--4 and Tables 3--5, the results obtained
depend strongly on the adopted solar velocity relative
to the local standard of rest.

The following estimates were obtained from 18 Cepheids with proper motions from the Hipparcos (1997) and UCAC4 (Zacharias et al. 2012) catalogues in Bobylev (2013): $R_0=7.64\pm0.32$~kpc and $V_0=217\pm11$~km s$^{-1}$. In this case, the parameters
from Sch\"onrich et al. (2010) for the solar velocity (5) were used and the influence of the spiral density wave was taken into account. Thus, this result should be
compared with the data in Table 3. As a result, we
have good agreement between the estimates, but in
this paper we used a larger number of Cepheids and, therefore, we obtained considerably smaller errors $\varepsilon_{R_0}$ and $\varepsilon_{V_0}$.

The history of the determination of the solar velocity
relative to the local standard of rest is very
dramatic. For example, based on data from the Hipparcos
catalogue, Dehnen and Binney (1998) found
$(U,V,W)_\odot=(10.00,5.25,7.17)$~km s$^{-1}$. If we repeat the approach of Table 4 for $r>3.0$~kpc with these velocities, then we will obtain fairly small values,
$R_0=7.64\pm0.21$~kpc and $V_0=240\pm8$~km s$^{-1}$. The
parameters (5) found by Sch\"onrich et al. (2010) are
more balanced. At present, they are widely used in kinematic studies.

The parameters of the standard solar apex $(U,V,W)_\odot=(10.3,15.3,7.7)$~km s$^{-1}$ were determined from a small sample of bright stars and were fixed as the solar motion with a velocity of 20~km s$^{-1}$ in the direction ($\alpha,\delta)=(18^h,30^\circ$). However, these
parameters are widely used by radio astronomers to
maintain the continuity of the results obtained by
them since the 1950s.

In Bobylev and Bajkova (2014) we showed that
there is an influence of the Galactic spiral density
wave on the solar velocity relative to the local standard
of rest determined from young disk stars. Therefore,
in our opinion, the parameters (6) take into account
the solar velocity relative to the local standard
of rest at the point $R=R_0$ more rigorously, i.e., they
take into account the solar velocity perturbed by the
spiral density wave at this point. Therefore, the parameters
given in Tables 4 and 5 are more interesting
than those in Tables 2 and 3.

In the Introduction we gave the estimates of $R_0$ obtained as a mean from the analysis of numerous individual determinations. Here we will note several
most recent individual estimates. Leung et al. (2023)
obtained an estimate of $R_0=8.23\pm0.12$~kpc from
the kinematics of stars in the central Galactic bar
using data from the APOGEE DR17 (Apache Point
Observatory Galactic Evolution Experiment, Blanton
et al. 2017) and Gaia EDR3 catalogues supplemented by spectrophotometric distances. Hey et al. (2023) found $R_0=8108\pm106_{\rm stat.} \pm93_{\rm sys.}$~pc
using $\sim$190\,000 semiregular variables in the Galactic bulge.

It is interesting to note the paper by Gordon et al. (2023), where new absolute VLBI measurements of the radio source Sgr~A* within the third
International Celestial Reference Frame (ICRF~3)
are presented. The observations were carried out
at 52 epochs with VLBA at 24 GHz in the period
2006--2022. Based on the measured proper motion
of Sgr~A* in Galactic longitude, these authors estimated
$V_0=248.0\pm2.8$~km s$^{-1}$ (at the specified $R_0=8.178\pm0.022$~kpc).

 \section*{CONCLUSIONS}
We obtained new estimates of the Galactocentric distance of the Sun $R_0$ and the Galactic rotation velocity at the solar distance $V_0$. They were found
by a peculiar method based on the analysis of objects
located near the solar circle. We used both the method
in the modification of Sofue et al. (2011) and the
extended method with an additional allowance for the
influence of the spiral density wave. The extended
method was proposed previously by Bobylev (2013)
when analyzing small samples of classical Cepheids
and star-forming regions.

In this paper we used a sample of classical Cepheids from the catalogue by Skowron et al. (2019), where the distances to them were determined from the period--luminosity relation with a mean random error of about 5\%. The proper motions and line-of-sight
velocities of the Cepheids were taken from the
Gaia DR3 catalogue. We showed previously (Bobylev
et al. 2021) that the Cepheid scale of Skowron
et al. (2019) should be slightly extended. Therefore,
in this paper the distances to the Cepheids from this
catalogue were increased by 10\%. For our analysis we
selected $\sim$200 Cepheids with a mean age of 87 Myr
located in the distance range $7<R<9$~kpc.

The values of $R_0$ found lie within the range [7.8--8.3] kpc, depending on the heliocentric distance of the sample stars, on the adopted solar velocity relative to
the local standard of rest, and on whether or not the perturbations caused by the Galactic spiral density wave are taken into account. The dispersion of the
$R_0$ estimates is about 2~kpc and the error of the mean is 0.2 kpc. Similarly, the values of $V_0$ lie within the range [240--270]~km s$^{-1}$, where the dispersion of the $V_0$ estimates is 70--90~km s$^{-1}$ and the error of the mean is 8~km s$^{-1}$.

We deem the estimates of $R_0=8.24\pm0.20$~kpc and $V_0=268\pm8$~km s$^{-1}$ that were obtained from a sample of 93 Cepheids by taking into account the
perturbations from the Galactic spiral density wave to be the best ones. These Cepheids are farther than 3 kpc from the Sun. 

 \subsubsection*{CONFLICT OF INTEREST}
As author of this work, I declare that I have no
conflicts of interest.

 \subsubsection*{REFERENCES}
 {\small

\quad~~1. R. Abuter, A. Amorim, M. Baub\"ock, et al. (GRAVITY Collab.), Astron. Astrophys. 625, L10 (2019).

2. R. Abuter, A. Amorim, M. Baub\"ock, et al. (GRAVITY Collab.), Astron. Astrophys. 647, A59 (2021).

3. R. I. Anderson, H. Saio, S. Ekstr\"om, C. Georgy, and G. Meynet, Astron. Astrophys. 591, A8 (2016).

4. R. A. Benjamin, E. Churchwell, B. L. Babler, T. M. Bania, D. P. Clemens, M. Cohen, J. M. Dickey, R. Indebetouw, et al., Publ. Astron. Soc. Pacif. 115, 953 (2003).

5. J. Bland-Hawthorn and O. Gerhard, Ann. Rev. Astron. Astrophys. 54, 529 (2016).

6. M. R. Blanton, M. A. Bershady, B. Abolfathi, F. D. Albareti, C. Allende Prieto, A. Almeida, J. Alonso-Garcia, F. Anders, et al., Astron. J. 154, 28 (2017).

7. V. V. Bobylev, Astron. Lett. 39, 95 (2013).

8. V. V. Bobylev and A. T. Bajkova, Mon. Not. R. Astron. Soc. 441, 142 (2014).

9. V. V. Bobylev, A. T. Bajkova, A. S. Rastorguev, and M. V. Zabolotskikh, Mon. Not. R. Astron. Soc. 502, 4377 (2021).

10. V. V. Bobylev and A. T. Bajkova, Astron. Rep. 65, 498 (2021).

11. V. V. Bobylev and A. T. Bajkova, Res. Astron. Astrophys. 23, 045001 (2023).

12. A. G. A. Brown, A. Vallenari, T. Prusti, et al. (Gaia Collab.), Astron. Astrophys. 616, 1 (2018).

13. A. G. A. Brown, A. Vallenari, T. Prusti, et al. (Gaia Collab.), Astron. Astrophys. 649, 1 (2021).

14. X. Chen, S. Wang, L. Deng, R. de Grijs, and M. Yang, Astrophys. J. Suppl. Ser., 237, 28 (2018).

15. W. Dehnen and J. J. Binney, Mon. Not. R. Astron. Soc. 298, 387 (1998).

16. A.-C. Eilers, D. W. Hogg, H.-W. Rix, and M. K. Ness, Astrophys. J. 871, 120 (2021).

17. D. Gordon, A. de Witt, and C. S. Jacobs, Astron. J. 165, 49 (2023).

18. D. R. Hey, D. Huber, B. J. Shappee, J. Bland-Hawthorn, Th. Tepper-Garcia, R. Sanderson, S. Chakrabarti, N. Saunders, et al., arXiv: 2305.19319 (2023).

19. H. S. Leavitt, Ann. Harvard College Observ. 60, 87 (1908).

20. H. S. Leavitt and E. C. Pickering, Harvard College Observ. Circ. 173, 1 (1912).

21. H. W. Leung, J. Bovy, and J. T. Mackereth, Mon. Not. R. Astron. Soc. 519, 948 (2023).

22. C. C. Lin and F. H. Shu, Astrophys. J. 140, 646 (1964).

23. B. F. Madore, Astrophys. J. 253, 575 (1982).

24. Z. Malkin, in Advancing the Physics of Cosmic Distances, Proc. IAU Symp. No. 289, 2012, Ed. R. de Grijs and G. Bono (2013).

25. P. Mr\'oz, A. Udalski, D. M. Skowron, J. Skowron, I. Soszynski, P. Pietrukowicz, M. K. Szymanski, R. Poleski, S. Kozlowski, and K. Ulaczyk, Astrophys. J. 870, L10 (2019).

26. I. I. Nikiforov, in Order and Chaos in Stellar and Planetary Systems, Proc. Conf., August 17--24, 2003, St. Petersburg, Ed. by G. G. Byrd, K. V. Kholshevnikov, A. A. Myllari, I. I. Nikiforov and V. V. Orlov, ASP Conf. Proc. 316,
199 (2004).

27. T. Prusti, J. H. J. de Bruijne, A. G. A. Brown, et al. (Gaia Collab.), Astron. Astrophys. 595, 1 (2016).

28. M. J. Reid, Ann. Rev. Astron. Astrophys. 31, 345 (1993).

29. M. J. Reid, K. M. Menten, X. W. Zheng, A. Brunthaler, and Y. Xu, Astrophys. J. 705, 1548 (2009).

30. V. Ripepi, R. Molinaro, I. Musella, M. Marconi, S. Leccia, and L. Eyer, Astron. Astrophys. 625, 14 (2019).

31. P. L. Schechter, I.M. Avruch, J. A. R. Caldwell, and M. J. Keane, Astron. J. 104, 1930 (1992).

32. R. Sch\"onrich, J. Binney, and W. Dehnen, Mon. Not. R. Astron. Soc. 403, 1829 (2010).

33. D. M. Skowron, J. Skowron, P. Mr\'oz, A. Udalski, P. Pietrukowicz, I. Soszynski, M. Szymanski, R. Poleski, et al., Science (Washington, DC, U. S.) 365, 478 (2019).

34. Y. Sofue, T. Nagayama, M. Matsui, and A. Nakagawa, Publ. Astron. Soc. Jpn. 63, 867 (2011).

35. The HIPPARCOS and Tycho Catalogues, ESA SP--1200 (1997).

36. A. Udalski, M. K. Szyma\'nski, and G. Szyma\'nski, Acta Astron. 65, 1 (2015).

37. A. Vallenari, A. G. A. Brown, T. Prusti, et al. (Gaia Collab.), arXiv: 2208.0021 (2022).

38. S. Wang, X. Chen, R. de Grijs, and L. Deng, Astrophys. J. 852, 78 (2018).

39. N. Zacharias, C. T. Finch, T. M. Girard, et al., Strasbourg Catalogue No. I/322 (2012).
 }
 \end{document}